\documentclass[11pt]{article}
\pdfoutput=1
\usepackage[utf8]{inputenc}
\usepackage{braket,slashed,bm}
\usepackage{jheppub} \usepackage{braket,slashed,bm}
\usepackage{array,multirow}
\usepackage{amsmath}
\usepackage[normalem]{ulem}
\usepackage[T1]{fontenc}
\usepackage{mathrsfs}
\usepackage{tikz}

\usepackage{subcaption}
\usetikzlibrary{arrows,decorations.pathmorphing,backgrounds,positioning,fit,petri,automata,shadows,calendar,mindmap,
decorations.markings,calc}
\usepackage{lscape}
\usepackage{hyperref}
\usepackage{graphicx}
\usepackage{booktabs}
\usepackage{floatrow}
\newfloatcommand{capbtabbox}{table}[][\FBwidth]
\usepackage{feynmp-auto}
\def\beq{\begin{equation}}
\def\eeq{\end{equation}}
\def\bea{\begin{eqnarray}}
\def\eea{\end{eqnarray}}
\def\nn{\nonumber \\}
\def\hyp{\mathsf{y}}

\title{On expansions in neutrino effective field theory}
\author{
Gitte Elgaard-Clausen and Michael Trott\\
Niels Bohr International Academy and Discovery Centre, Niels Bohr Institute,
University of Copenhagen, Blegdamsvej 17, DK-2100 Copenhagen, Denmark}
\abstract{We match the seesaw model for generating neutrino masses onto the Standard Model Effective Field Theory (SMEFT).
We perform this matching at tree level  up to dimension seven in the operator expansion.
We explain how some of the perturbations of the neutrino mass matrix due to operators of mass dimension greater than five are tied
to integrating out the heavy Majorana mass eigenstates in sequence. We demonstrate that the low energy limit of seesaw models
are well described by the SMEFT, particularly when constructed using a flavour space expansion.
Flavour space expansions of seesaw models are of interest as the coupling of the heavy states to the Standard Model, that are integrated out to generate neutrino masses,
are through flavour space vectors $\in \mathbb{C}^3$. We point out that neutrino phenomenology can be systematically developed as
a perturbation around the unknown eigenvectors diagonalizing the charged lepton mass matrix using the fact that these eigenvectors also form a
basis of $\mathbb{C}^3$. This point holds in seesaw models and can also be applied to other models of neutrino mass generation to develop systematic expansions. We develop the algebra for this flavour space and discuss some phenomenology to illustrate this approach.}
\begin{document}
\maketitle

\section{Introduction} \label{sec:intro}
Recently, there has been an escalation of theoretical efforts treating the Standard Model (SM) as a consistent low energy limit of a more fundamental theory.
This is a natural result of the discovery of a dominantly $J^P = 0^+$ Higgs like boson at the Large Hadron Collider (LHC), and increased experimental indications
that there is a mass gap between the electroweak scale ($v \sim 246$~{\rm GeV}) and any scale of new physics. It is reasonable to assume that the SM Lagrangian terms are the leading terms in
the Standard Model Effective Field Theory (SMEFT) operator expansion \cite{Weinberg:1979sa,Wilczek:1979hc,Buchmuller:1985jz,Grzadkowski:2010es,Abbott:1980zj,Lehman:2015via,Lehman:2014jma,Lehman:2015coa,Henning:2015alf}.

Despite the power of the SMEFT formalism and recent systematic developments, there is little direct experimental evidence that higher dimensional operators supplementing the SM
have non-vanishing Wilson coefficients. One exception is arguably supplied by the Wilson coefficient of the dimension five operator given by
 \cite{Weinberg:1979sa,Wilczek:1979hc},
\beq\label{weinberg}
\mathcal{Q}^{\beta \, \kappa}_5 = \left(\overline{\ell^{c, \beta}_{L}} \, \tilde{H}^\star\right) \left(\tilde{H}^\dagger \, \ell_{L}^{\kappa}\right).
\eeq
This operator\footnote{The $c$ superscript in Eqn.~\ref{weinberg} corresponds to a charge conjugated Dirac four component spinor defined as $\psi^c  = C \overline{\psi}^T$ with $C= - i \gamma_2 \, \gamma_0$ in the chiral basis we employ. The star superscript is reserved for the complex conjugation operation that is applied to scalar and vector quantities. Chiral projection and charge conjugation do not commute (see the discussion in Ref.~\cite{Willenbrock:2004hu} for a review).
$\ell_L^c$ denotes the doublet lepton field that is chirally projected and subsequently charge conjugated.} violates global Lepton number, $\rm U(1)_L$, which is accidentally preserved in the SMEFT operators of mass dimension less than or equal to four \cite{deGouvea:2014lva,Kobach:2016ami}.
For this reason it is generally neglected in LHC studies of the SMEFT.  A nonzero
value of this operator's Wilson coefficient leads to Majorana neutrino mass terms, which are not present in the minimal SM Lagrangian. Masses for neutrinos are now strongly experimentally supported \cite{Olive:2016xmw} which makes it appealing
to obtain a nonzero Wilson coefficient for this operator. Arguably the simplest way  to generate this Wilson coefficient is to directly integrate out heavy singlet fields extending the SM (here denoted $N_{p}$) using a seesaw mechanism
of neutrino mass generation \cite{Minkowski:1977sc,GellMann:1980vs,Yanagida:1979as,Mohapatra:1979ia}. A seesaw mechanism for neutrino mass provides an explanation of the smallness of neutrino masses
due to a hierarchy of scales. Such an extension of the SM is well described by an effective field theory approach for the same reason.

In this paper, we systematically develop the SMEFT implementation and matching of the seesaw model, integrating out the heavy $N_{p}$ states
assuming a renormalizable ultraviolet (UV) extension to the SM. We examine the effect of higher dimensional operators in the SMEFT operator expansion, beyond the Weinberg operator, on the low energy neutrino mass matrix that results.
We find by explicit calculation the tree level matching contributions to the SMEFT dimension seven operators.

The seesaw model has been studied many times in the past in an EFT context, see
Refs.\cite{Broncano:2002rw,Gavela:2009cd,Gavela:2008ra,Abada:2007ux,Broncano:2003fq,Bonnet:2012kz,Bonnet:2009ej,delAguila:2008ir,delAguila:2012nu,Bhattacharya:2015vja,Angel:2012ug}.
Our results go beyond past work by reporting the complete matching for three generations of heavy singlet fields integrated out in sequence in the seesaw model for the first time
up to dimension seven.
Simultaneously we incorporate into this implementation of neutrino EFT the flavour space expansion of neutrino phenomenology previously developed in Ref.\cite{Grinstein:2012yu}. We discuss how flavour space expansions can be used to relate the neutrino mass spectrum to the Pontecorvo-Maki-Nakagawa-Sakata (PMNS) matrix \cite{Pontecorvo:1957cp,Maki:1962mu} phenomenology. We extend the results of  Ref.\cite{Grinstein:2012yu} using the fact that
one can perturb neutrino phenomenology about the eigenvectors diagonalizing the lepton mass matrix in a general way, simply treating these vectors
as an unknown basis of $\mathbb{C}^3$. We demonstrate the utility of the systematic expansion that can be constructed using this technique with a simple phenomenological example.

The method developed here can be used to study the growing data set on neutrino phenomenology.
This can be done in a systematically improvable manner, using well defined expansions, in an effective field theory approach. This formalism is sufficiently general that it can accommodate flavour symmetries assumed in the UV sector,
but is not limited to any such flavour symmetry requirement. This approach can also be extended to other UV models of neutrino mass generation in a straightforward manner.
\section{Full theory for a minimal seesaw scenario}
We consider the full theory Lagrangian\footnote{We acknowledge that explicit mass scales are introduced without a dynamical origin in this "full theory" --  {\it castigat ridendo mores}.} as given by $\mathcal{L}= \mathcal{L}_{SM} +  \mathcal{L}_{N_p}$.
To fix our notation we define the SM Lagrangian ($\mathcal{L}_{SM}$) as
\bea\label{SML}
\mathcal{L}_{SM} &=& - \frac{1}{4} \left(G^A_{\mu \nu} G^{A\mu \nu} + W^I_{\mu \nu} W^{I \mu \nu} +B_{\mu \nu} B^{\mu \nu}\right)+ (D_\mu H)^\dagger(D^\mu H) + \sum_{\psi} \bar{\psi} i \slashed{D} \psi  \\
&-& \Big( H^{\dagger j} \, \bar{d}_R \, Y_d \, Q_{L j} +\tilde{H}^{\dagger j} \, \bar{u}_R \, Y_u \, Q_{L j} + H^{\dagger j} \,\bar{e}_R \, Y_e \, \ell_{L_j} + h.c. \Big) - \lambda \big(H^\dagger H- \frac{1}{2}v^2 \big)^2. \nonumber \label{SMNoN}
\eea
Here the fermion fields summed over are $\psi = \{Q_L,u_R, d_R, \ell_L,e_R\}$ and the fields in $\mathcal{L}_{SM}$ are written in the weak eigenstate basis.
$\tilde{H}_j= \epsilon_{jk}H^{\dagger k}$, where $\epsilon_{12}=1$ and $\epsilon_{jk}=-\epsilon_{kj}$, $j,k=\{1,2\}$. $H^j$ is the Higgs field of the SM with labeled $\rm SU(2)_L$ components, conventionally indicated with Roman letters, usually $\{j,k,l,m,n\}$ in this work. At times we suppress the explicit  $\rm SU(2)_L$  indicies on the $\epsilon_{ij}$ tensor.
The Higgs mass is  defined as $m_H^2 = 2 \, \lambda \, v^2$.
The fermion mass matrices are $M_{u,d,e}=Y_{u,d,e}\, v /\sqrt 2$. The $M_{u,d,e}$ and Yukawa matrices $Y_{u,d,e}$ are complex matrices in flavour space.
The gauge covariant derivative is defined as
\bea
D_\mu = \partial_\mu + i g_3 T^A A^A_\mu + i g_2  t^I W^I_\mu + i g_1 \hyp B_\mu,
\eea
where $T^A$ are the $\rm SU(3)$ generators,  $t^I=\tau^I/2$ are the $\rm SU(2)$ generators, and $\hyp$ is the $\rm U(1)$ hypercharge generator.  Flavour indicies are suppressed in Eqn.~\ref{SML},
restoring the flavour indicies one has for example: $H^{\dagger j} \overline d\, Y_d\, q_{j} \rightarrow H^{\dagger j} \overline d_p\, [Y_d]_{pr}\, q_{rj}$
where the flavor indicies (conventionally $p,r,s,t$) are summed over $\{1,2,3\}$ for the three generations.

The extension of the SM Lagrangian from a right handed singlet field $N_R$ with
vanishing $\rm SU(3) \times SU(2)_L \times U(1)_Y$ charge is well known. Such fields can have Majorana mass terms \cite{Majorana:1937vz} of the form
\bea
\overline{N_{R,p}^c} \, M_{pr} \, N_{R,r} + \overline{N_{R,p}} \, M^\star_{pr} \, N_{R,r}^c,
\eea
where the charge conjugate of $N_R$ is $N_{R}^c$.
Following Ref.~\cite{Broncano:2002rw}, we define a field satisfying the Majorana condition $N_p = N_p^c$
in its mass eigenstate basis, with all Majorana phases $\theta_p$ for each real mass eigenstate shifted into the effective
couplings \cite{Broncano:2002rw},
\bea
N_p = e^{i \theta_p/2} \, N_{R,p} + e^{- i \theta_p/2} \, (N_{R,p})^{c}.
\eea
The corresponding Lagrangian is defined as
\bea\label{basicL}
2 \, \mathcal{L}_{N_p}=  \overline{N_p} (i\slashed{\partial} - m_{p})N_p - \overline{\ell_{L}^\beta} \tilde{H} \omega^{p,\dagger}_\beta  N_p -  \overline{\ell_{L}^{c \beta}} \tilde{H}^* \, \omega^{p,T}_\beta N_p - \overline{N_p} \, \omega^{p,*}_\beta \tilde{H}^T \ell_{L}^{c \beta}  - \overline{N_p}\, \omega^p_\beta \tilde{H}^\dagger \ell_{L}^\beta.  \label{LNa}
\eea
The $\omega^p_\beta = \{x_\beta,y_\beta,z_\beta\}$ are each complex vectors in flavour space that have absorbed the Majorana phases. The invariants constructed from these vectors will allow a flavour space expansion as we discuss below. $N_p$ is a four component spinor satisfying the Majorana condition, not a two component Weyl spinor.
We use greek letters such as $\beta, \kappa$ for the label of a flavour vector in the heavy singlet field mass eigenbasis.
\subsection{Equations of motion of the seesaw theory}
We integrate out each $N_p$ in sequence, and utilize the Equation of Motion (EOM) to reduce to an operator basis. The EOM include the $N_p$ states still present in the spectrum. The relevant modifications of the SM EOM are
\bea
D^2 H_j  &=& \lambda \left(v^2 - 2 (H^\dagger H) \right)H_j - \overline Q_L^k\, Y_u^\dagger\, u_R \, \epsilon_{kj} - \overline d_R\, Y_d \, Q_{L j} - \overline e_R \, Y_e\,  \ell_{L j}, \nonumber \\
&-&  \frac{1}{2}  \overline{\ell_{L}}^{k \beta }\epsilon_{kj}  \, (\omega^p_\beta)^\dagger  N_p  + \frac{1}{2} \overline{N_p} \, \omega^{p,*}_\beta \epsilon_{jk} \ell_{L}^{c, \, k \, \beta},
 \label{EOMH}
\eea
and
\begin{align}
i \slashed{D} (\ell^j_{L})_\beta=  (Y_e^\dagger)_{\beta s} e_{R}^s H^j + \tilde{H}^j \omega^{p,\dagger}_\beta N_p
- c^\star_{\kappa \beta} \, \tilde{H}^j (\tilde{H}^T \, \ell_L^{c, \kappa})
.    
\label{EOMell}
\end{align}
Note the last term in the EOM due to varying the fields in the $\mathcal{L}_5$ operator in the SMEFT.\footnote{Note that a series of
$1/m_i^n$ terms also exist correcting the right hand side of Eqn.~\ref{EOMH} in the SMEFT, including correction due to $\mathcal{L}_5$, but these terms are
relatively further supressed for the final results.
Such corrections due to the Higgs EOM do not lead to $\mathcal{L}^{(7)}$ matching corrections as
the dimensionality of the fields in the $\mathcal{L}^{(6)}$ matching only allow one derivative insertion, while the Higgs EOM has two derivatives.}
Finally the EOM for the $N_{p}$ are
\bea\label{NEOM}
 \slashed{\partial} N_p =& -i\Big( m_{p} \, N_p + w^{p,*}_\beta \tilde{H}^T \ell_L^{c \beta} + w^{p}_\beta \tilde{H}^\dagger \ell_L^{ \beta} \Big).
\eea
The usage of the EOM consistently drops a $N_p$ field when it is integrated out of the theory.
\section{Matching the Seesaw to the SMEFT} \label{sec:SMEFTintro}
Integrating out the $N_p$ we match onto the SMEFT. The SMEFT is  defined as
the sum of $\rm SU(3)_C \times SU(2)_L \times U(1)_Y$ invariant higher dimensional operators built out of SM fields
\bea
\mathcal{L}_{SMEFT} = \mathcal{L}_{SM} + \mathcal{L}^{(5)} + \mathcal{L}^{(6)} + \mathcal{L}^{(7)} + ..., \quad \quad \mathcal{L}^{(k)}= \sum_{\alpha = 1}^{n_k} \frac{C_\alpha^{(k)}}{\Lambda^{k-4}} \mathcal{Q}_\alpha^{(k)} \hspace{0.25cm} \text{ for $k > 4$. }
\eea
Here $\mathcal{L}^{(k)}$ contains the dimension $k$ operators $\mathcal{Q}_\alpha^{(k)}$. The number of non redundant operators in $\mathcal{L}^{(5)}$, $\mathcal{L}^{(6)}$, $\mathcal{L}^{(7)}$ and $\mathcal{L}^{(8)}$ is known \cite{Buchmuller:1985jz,Grzadkowski:2010es,Weinberg:1979sa,Abbott:1980zj,Lehman:2015via,Lehman:2014jma,Lehman:2015coa,Henning:2015alf}. Past works on $\mathcal{L}^{(7)}$ operator bases particularly relevant to this study are Refs.~\cite{Lehman:2014jma,Lehman:2015coa,Henning:2015alf,Liao:2016hru,Bhattacharya:2015vja,Liao:2016qyd}.

\begin{figure}[h!]
\includegraphics[width=0.45\textwidth]{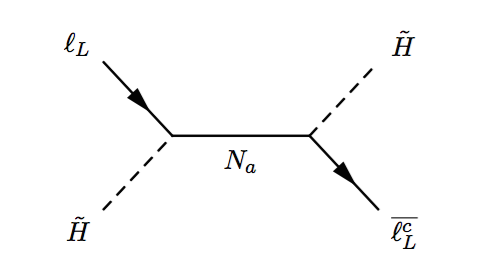}
\caption{\label{fig:IntVer} Tree level exchange expanded out to match onto $\mathcal{L}^{(5)}$, $\mathcal{L}^{(6)}$, $\mathcal{L}^{(7)} \cdots$.}
\end{figure}

Matching onto the SMEFT is defined by requiring that the Wilson coefficients in the higher dimensional operators reproduce the low energy, or infrared (IR),
limit of the full theory.  For example, consider the IR limit where $s^2 \ll m_p^2$ for $N_p$ carrying four momenta $s^\mu$ as illustrated in Fig.~\ref{fig:IntVer}.
The $N_p$ $s$-channel
propagator is expanded in this limit as
\bea\label{propexpansion}
(\slashed{s}+m_{p})\frac{-1}{m_{p}^2} \Big( \frac{1}{1-s^2/m_{p}^2} \Big) =  -\frac{1}{m_{p}} - \frac{\slashed{s}}{m_{p}^2} - \frac{s^2}{m_{p}^3} + \cdots
\eea
Note that we adopt a conventional normalization of the Wilson coefficient of the dimension five operator of the form
\bea
\mathcal{L}^{(5)} = \frac{c_{\beta \kappa}}{2}  \, \mathcal{Q}^{\beta \kappa}_5 + {\rm h.c.}
\eea
\subsection{$\mathcal{L}^{(5)}$ matching}
Integrating out the heaviest $N_p$ state, denoted $N_1$, the matching onto the leading $\mathcal{L}^{(5)}$ operators is given by
\bea
\mathcal{L}^{(5)} = \frac{(x_\beta)^T \, x_\kappa}{2 \, m_1} \,  \mathcal{Q}^{\beta \, \kappa}_5 + h.c.
\eea
The matrix $ (x_\beta)^T \, x_\kappa/m_1$ is complex with only one eigenvalue, as
only $N_1$ was integrated out coupled to the complex flavour vector $x_\beta$. The notation $x_\beta^T \, x_\kappa$ is an outer product of the complex vectors.
Integrating out the remaining two lighter $N_p$ states in sequence gives
\bea
\mathcal{L}^{(5)} = \frac{c_{\beta \, \kappa}}{2} \,  \mathcal{Q}^{\beta \, \kappa}_5 \,  + h.c.
\eea
where $c_{\beta \, \kappa} =  (\omega^p_\beta)^T \, \omega^p_\kappa/m_p$ and the flavour index $p$ is summed over.
Contracting the $\rm SU(2)_L$ indicies of $\mathcal{Q}_5$ and taking a matrix element  where the Higgs field is taken as the background field value gives
\bea
\langle c_{\beta \, \kappa} \, \mathcal{Q}^{\beta \, \kappa}_5 \rangle  &=& \frac{v^2 \, c_{\beta \, \kappa}}{2} \, \overline{\nu_L^{c \, \beta}} \, \nu_L^\kappa.
\eea
We define the mass eigenstate neutrino fields with prime superscripts. These field are related by the unitary rotation matricies (denoted $\mathcal{U}$) to the weak eigenstates used
so far by
\bea
\nu_L^p = \mathcal{U}(\nu,L)^{p}_r \,  \nu_{L}^{' r},
\eea
Changing to the mass eigenstate basis we find
\bea
\langle c_{\beta \, \kappa}\, \mathcal{Q}^{\beta \, \kappa}_5 \rangle  &=& - \frac{v^2}{2} \, \left[\mathcal{U}^T(\nu,L)^{\beta}_p \, c_{\beta \, \kappa}  \, \mathcal{U}(\nu,L)^{\kappa}_r\right] (\nu'_L)^{T p} \, \epsilon \, (\nu'_L)^{r},
\eea
where  $\mathcal{U}^T(\nu,L)^{\beta}_p \, c_{\beta \, \kappa}  \, \mathcal{U}(\nu,L)^{\kappa}_r \equiv - {\rm diag}\{C^1,C^2,C^3 \}_{pr}$. The physical low energy neutrino masses $m_{\nu}^p$
at leading order in the SMEFT expansion in $v/m_{p}$ are then given as\footnote{The overall sign in the Majorana mass term is linked to the phase convention choice on $C$.}
\bea
m_{\nu}^p= \frac{v^2}{2} \, C^p.
\eea
\subsection{$\mathcal{L}^{(6)}$ matching}
The $\mathcal{L}^{(6)}$ matching follows directly and we find
\bea\label{dim6match}
\mathcal{L}^{(6)} &=& \frac{(\omega^p_\beta)^\dagger \, \omega^p_\kappa}{2 \, m_p^2} \,
\left(\mathcal{Q}^{(1)}_{\substack{H \ell \\ \beta \kappa}} - \mathcal{Q}^{(3)}_{\substack{H \ell \\ \beta \kappa}}\right).
\eea
The operators $Q^{(1)}_{H \ell},Q^{(3)}_{H \ell}$ each with flavour indicies $\beta \kappa$ are defined as in Ref.~\cite{Grzadkowski:2010es} with the notation $\phi$ exchanged for
$H$ for the Higgs field.\footnote{Explicitly these operators are given by $\mathcal{Q}^{(1)}_{\substack{H \ell \\ \beta \kappa}} = H^\dagger \, i\overleftrightarrow D_\mu H
\ell_\beta \gamma^\mu \ell_\kappa$ and
$\mathcal{Q}^{(3)}_{\substack{H \ell \\ \beta \kappa}} = H^\dagger \, i\overleftrightarrow D^I_\mu H
\ell_\beta \gamma^\mu  \tau_I \ell_\kappa$}
Here we have reduced the operators to the Warsaw basis form using the EOM and combining terms into Hermitian derivatives
defined as $H^\dagger \, i\overleftrightarrow D_\beta H = i H^\dagger (D_\beta H) - i (D_\beta H)^\dagger H$
and $H^\dagger \, i\overleftrightarrow D_\beta^I H = i H^\dagger \tau^I (D_\beta H) - i (D_\beta H)^\dagger\tau^I H$.
We have used the fact that Hermitian operators generate real eigenvalues, and hence the matching coefficient in Eqn.~\ref{dim6match} is real.
The derivative on the lepton doublet field has been reduced out using the EOM and
using the fact that $\tilde{H}^\dagger H = 0$. For previous results on dimension six matching comparable to the terms
in Eqn.~\ref{dim6match} see Ref.~\cite{Broncano:2002rw,Broncano:2003fq,Broncano:2004tz,Abada:2007ux}. Our results
are distinct from past works in the $\rm SU(2)_L$ field dependence. As Eqn.\ref{EOMH} contains the $N_p$ fields still in the spectrum when integrating out
the heavy Majorana mass eigenstates in sequence, the following terms are also generated.
Integrating out $N_1$ gives
\bea\label{EOMdim6}
\mathcal{L}^{(6),N_1}_{N_{2,3}} &\supseteq& \frac{{\rm Re} \left[x_\beta^\dagger \, x^\star \cdot y^\dagger \right]}{2 \, m_1^2} \left(\mathcal{Q}^\beta_{N_2}- \mathcal{Q}^{\star,\beta}_{N_2}\right)+
\frac{i \, {\rm Im} \left[x_\beta^\dagger \, x^\star \cdot y^\dagger \right]}{2 \, m_1^2} \left(\mathcal{Q}^\beta_{N_2}+ \mathcal{Q}^{\star,\beta}_{N_2}\right), \nonumber \\
&+& \frac{{\rm Re} \left[x_\beta^\dagger \, x^\star \cdot z^\dagger \right]}{2 \, m_1^2} \left(\mathcal{Q}^\beta_{N_3}- \mathcal{Q}^{\star,\beta}_{N_3}\right)+
\frac{i \, {\rm Im} \left[x_\beta^\dagger \, x^\star \cdot z^\dagger \right]}{2 \, m_1^2} \left(\mathcal{Q}^\beta_{N_3}+ \mathcal{Q}^{\star,\beta}_{N_3}\right),
\eea
integrating out $N_2$ gives
\bea\label{EOMdim62}
\mathcal{L}^{(6),N_2}_{N_{3}} &\supseteq&
 \frac{{\rm Re} \left[y_\beta^\dagger \, y^\star \cdot z^\dagger \right]}{2 \, m_2^2} \left(\mathcal{Q}^\beta_{N_3}- \mathcal{Q}^{\star,\beta}_{N_3}\right)+
\frac{i \, {\rm Im} \left[y_\beta^\dagger \, y^\star \cdot z^\dagger \right]}{2 \, m_2^2} \left(\mathcal{Q}^\beta_{N_3}+ \mathcal{Q}^{\star,\beta}_{N_3}\right),
\eea
where $\mathcal{Q}_{N_p}^\beta = (H^\dagger H) \, (\overline{\ell_L^\beta} \tilde{H}) \, N_p$. Here the notation $a \cdot b$ applied to complex flavour vectors $\{x,y,z\}$
is a Hermitian inner product, see the Appendix for details on the flavour space algebra.

Due to the presence of the Majorana mass scale in the EOM the following contributions to $\mathcal{L}_{N_{2,3}}^{(6)}$ are also present.
Integrating out $N_1$
\bea\label{L6N23}
\mathcal{L}^{(6),N_1}_{N_{2,3}} &\supseteq& \frac{(x_\beta)^T \, x^\star \cdot y^\dagger \, m_2}{2 \, m_1^3} \, \left[\overline{\ell_{L \, \beta}^{c}} \, \tilde{H}^\star \, N_2 \right]  \, (H^\dagger \, H)
+  \frac{(x_\beta)^T \, x^\star \cdot z^\dagger \, m_3}{2 \, m_1^3} \, \left[\overline{\ell_{L \, \beta}^{c}} \, \tilde{H}^\star \, N_3 \right]  \, (H^\dagger \, H),\nn
\eea
while integrating out $N_2$ gives
\bea\label{L6N232}
\mathcal{L}^{(6),N_2}_{N_{3}} &\supseteq& \frac{(y_\beta)^T \, y^\star \cdot z^\dagger \, m_3}{2 \, m_2^3} \, \left[\overline{\ell_{L \, \beta}^{c}} \, \tilde{H}^\star \, N_3 \right]  \, (H^\dagger \, H) + h.c.
\eea
\subsection{$\mathcal{L}^{(7)}$ matching}\label{directdim7match}
Dimension seven operators come about due to the expansion of a propagator, such as Eqn.~\ref{propexpansion}, to third order,
and from the contraction of the local contact operators present in $\mathcal{L}^{(6)}_{N_{2,3}}$  once $N_{1,2}$
are integrated out in time ordered products. We follow the approach in Refs.~\cite{Grzadkowski:2010es,Lehman:2014jma} of removing derivative
operators in the basis. We
define the short hand notation to aid in presenting the results
\bea
\tilde{C}^7_{\beta \, \kappa} = \sum_p \, \frac{(\omega^p_\beta)^T \, \omega^p_\kappa}{2 \, m_p^3}.
\eea

Using the Higgs EOM in Eqn.~\ref{EOMH} on the results of the tree level exchange of $N_p$ expanded to third order, one finds the terms
\bea\label{HEOMterms}
\mathcal{L}^{(7)} &\supseteq&- \frac{\lambda \, v^2 \,  \tilde{C}^7_{\beta \, \kappa}}{2} \,  \left(\overline{\ell_{L \, \beta}^{c}} \, \ell_{L \, \kappa} \right) \, H^2  + 2 \, \lambda \,  \tilde{C}^7_{\beta \, \kappa} \mathcal{Q}_{\ell H}
+ \frac{\lambda \, v^2 \, \tilde{C}^7_{\beta \, \kappa}}{2} \left(\overline{\ell_{L \, \beta}^{c}} \, \sigma^I \, \ell_{L \, \kappa} \right) \, H \sigma^I H  + h.c \nonumber \\
\eea
$\mathcal{Q}_{LH}$
and the remaining operator notation for $\mathcal{L}^{(7)}$ is defined in Ref.~\cite{Lehman:2014jma}.\footnote{The explicit operator
definitions for $\mathcal{L}^{(7)}$ are listed in the Appendix for completeness.}
Eqn.~\ref{HEOMterms}  vanishes when the Higgs takes on its background expectation value. This leads to a vanishing of the contributions to the low energy neutrino mass
matrix from this sum of terms. Applying the Higgs EOM and reducing the direct matching contributions into field strengths of the SM fields leads to the $\mathcal{L}^{(7)}$ operators\footnote{Note that the renormalizable weakly coupled seesaw model induces operators with field strengths in $\mathcal{L}^{(7)}$ at tree level. This is expected on general grounds
in well defined EFT's ~\cite{Jenkins:2013fya}.}
\bea\label{dimsevenops}
\mathcal{L}^{(7)} &\supseteq& - \tilde{C}^7_{\beta \, \kappa} Y_u^\dagger  \mathcal{Q}^{\kappa \, \beta}_{\ell \ell \bar{Q} u H}
- (\tilde{C}^7_{\kappa \, \beta} -\tilde{C}^7_{\beta \, \kappa}) \, Y_d \, \mathcal{Q}^{(1) \, \beta \, \kappa}_{\ell \ell Q \bar{d} H} -
\tilde{C}^7_{\beta \, \kappa} \, Y_d \, \mathcal{Q}^{(2)  \,  \beta \, \kappa}_{\ell \ell Q \bar{d} H}
+ \tilde{C}^7_{\beta \, \kappa} \, Y_e \,  \mathcal{Q}^{\kappa \beta}_{\ell \ell \ell \bar{e} H}, \nn
&+& g_1\, \hyp_\ell \, \tilde{C}^7_{\beta \, \kappa}  \, \mathcal{Q}^{\beta \, \kappa}_{\ell HB} + \frac{g_2 \, \tilde{C}^7_{\beta \, \kappa}}{2} \, \mathcal{Q}^{\beta \, \kappa}_{\ell HW}
- i \,\tilde{C}^7_{\beta \, \kappa} \, (Y_e^\dagger)_\kappa^\alpha  \, \mathcal{Q}^\beta_{\ell HDe_\alpha} +
 \frac{(x_\beta)^T \, x^\star \cdot y^\dagger \, y_\delta}{2 \, m_1^3} \, \mathcal{Q}_{\ell H}^{\beta \, \delta}, \nn
&+&  \frac{(x_\beta)^T \, x^\star \cdot z^\dagger \, z_\delta}{2 \, m_1^3} \, \mathcal{Q}_{\ell H}^{\beta \, \delta}
+  \frac{(y_\beta)^T \, y^\star \cdot z^\dagger \, z_\delta}{2 \, m_2^3} \, \mathcal{Q}_{\ell H}^{\beta \, \delta}
- 2 \, \tilde{C}^7_{\beta \, \kappa} \, \mathcal{Q}_{\ell HD}^{(2)} + h.c.
\eea
Here we have used Fierz relations and the EOM to reduce to this basis, utilizing Refs.~\cite{Jenkins:2013zja,Dreiner:2008tw,Liao:2016hru}.
It is also important to include the effect of $\mathcal{L}_5$ in determining the EOM
for the lepton fields, as this contribution leads to a matching contribution to $\mathcal{Q}_{\ell H}$ of the form
\bea\label{EOMeffectslepton}
\mathcal{L}^{(7)} &\supseteq&  - \left[\frac{x_\beta^T \, x_\kappa \, ||x||}{2 m_1^3} + \frac{y_\beta^T \, y_\kappa ||y||}{2 m_2^3} + \frac{z_\beta^T \, z_\kappa\, ||z||}{2 m_3^3}\right] \, \mathcal{Q}_{\ell H},\nn
&\,&- \left[\frac{y_\beta^T \, x_\kappa y \cdot x}{2 m_2^2 \, m_1} + \frac{z_\beta^T \, x_\kappa z \cdot x}{2 m_3^2 \, m_1} + \frac{z_\beta^T \, y_\kappa z \cdot y}{2 m_3^2 \, m_2} \right] \, \mathcal{Q}_{\ell H} + h.c.
\eea
This contribution perturbs the neturino mass matrix, as we discuss below.
The operators in $\mathcal{L}_{N_{2,3}}^{(7)}$ when $N_1$ is integrated out are given by
\bea
\mathcal{L}^{(7)}_{N_{2,3}}&\supseteq& \frac{(x_\beta)^T \, x_\kappa \, y^\dagger_\alpha}{8 \, m_1^3} \, \left[\overline{\ell_{L \, \beta}^{c}} \,  \ell_{L \,  \kappa} \right] \, \overline{\ell_L^\alpha} \, \tilde{H}^\star  \, N_2
+ \frac{(x_\beta)^T \, x_\kappa \, z^\dagger_\alpha}{8 \, m_1^3} \, \left[\overline{\ell_{L \, \beta}^{c}} \,  \ell_{L \,  \kappa} \right] \, \overline{\ell_L^\alpha} \, \tilde{H}^\star  \, N_3, \nn
&+& \frac{(x_\beta)^T \, x_\kappa \, y^\dagger_\alpha}{8 \, m_1^3} \, \left[\overline{\ell_{L \, \beta}^{c}}\, \sigma^I \,  \ell_{L \,  \kappa} \right] \,  \overline{\ell_L^\alpha} \, \sigma^I \, \tilde{H}^\star   \,  N_2
+ \frac{(x_\beta)^T \, x_\kappa \, z^\dagger_\alpha}{8 \, m_1^3} \, \left[\overline{\ell_{L \, \beta}^{c}} \, \sigma^I \,   \ell_{L \,  \kappa} \right] \,  \overline{\ell_L^\alpha} \, \sigma^I \, \tilde{H}^\star  \,  N_3, \nn
&+& \frac{(x_\beta)^T \, x_\kappa \, y^\star_\alpha}{8 \, m_1^3} \, \left[\overline{\ell_{L \, \beta}^{c}} \,  \ell_{L \,  \kappa} \right] \overline{N_2} \, \ell_{L \, \alpha}^{c} \, \tilde{H}^\star
+ \frac{(x_\beta)^T \, x_\kappa \, z^\star_\alpha}{8 \, m_1^3}\, \left[\overline{\ell_{L \, \beta}^{c}} \,  \ell_{L \,  \kappa} \right]
\overline{N_3} \, \ell_{L \, \alpha}^{c} \, \tilde{H}^\star, \\
&+& \frac{(x_\beta)^T \, x_\kappa \, y^\star_\alpha}{8 \, m_1^3} \, \left[\overline{\ell_{L \, \beta}^{c}} \,  \sigma^I \, \ell_{L \,  \kappa} \right]  \overline{N_2} \, \sigma^I \, \ell_{L \, \alpha}^{c} \, \tilde{H}^\star
+ \frac{(x_\beta)^T \, x_\kappa \, z^\star_\alpha}{8 \, m_1^3}\, \left[\overline{\ell_{L \, \beta}^{c}} \, \sigma^I \,  \ell_{L \,  \kappa} \right]
 \overline{N_3} \, \sigma^I \, \ell_{L \, \alpha}^{c} \, \tilde{H}^\star,\nonumber \\
&+& \frac{i (x_\beta)^T x^\star \cdot y^\dagger}{2 \, m_1^3}  \left[\overline{\ell_{L \beta}^{c}}  \gamma_\mu N_2 \right]  \tilde{H}^\star  (H D_\mu H^\dagger)
+ \frac{i (x_\beta)^T  x^\star \cdot z^\dagger}{2 \, m_1^3}  \left[\overline{\ell_{L  \beta}^{c}}  \gamma_\mu N_3 \right] \tilde{H}^\star  (H  D_\mu H^\dagger)  + h.c. \nonumber
\eea
In addition, when $N_2$ is integrated out in sequence the additional matching contributions to the operators involving $N_3$ are
\bea
\mathcal{L}^{(7)}_{N_{3}} &\supseteq& \frac{(y_\beta)^T \, y_\kappa \, z^\dagger_\alpha}{8 \, m_2^3} \, \left[\overline{\ell_{L \, \beta}^{c}} \,   \ell_{L \,  \kappa} \right] \, \overline{\ell_L^\alpha} \, \tilde{H}^\star  \, N_3
 + \frac{(y_\beta)^T \, y_\kappa \, z^\star}{8 \, m_2^3} \, \left[\overline{\ell_{L \, \beta}^{c}} \,  \ell_{L \,  \kappa} \right] \overline{N_3} \,  \ell_{L \, \alpha}^{c} \, \tilde{H}^\star, \nn
&+&\frac{(y_\beta)^T \, y_\kappa \,  z^\dagger_\alpha}{8 \, m_2^3} \, \left[\overline{\ell_{L \, \beta}^{c}} \,  \sigma^I \,  \ell_{L \,  \kappa} \right] \overline{\ell_L^\alpha} \, \sigma^I \, \tilde{H}^\star   \,  N_3  + \frac{(y_\beta)^T \, y_\kappa}{8 \, m_2^3} \, \left[\overline{\ell_{L \, \beta}^{c}} \, \sigma^I \, \ell_{L \,  \kappa} \right] \overline{N_3} \, \sigma^I \, \ell_{L \, \alpha}^{c} \, \tilde{H}^\star, \nonumber \\
&+& \frac{i \, (y_\beta)^T \, y^\star \cdot z^\dagger}{2 \, m_2^3} \, \left[\overline{\ell_{L \, \beta}^{c}} \, \gamma_\mu N_3 \right] \, \tilde{H}^\star \, (H \, D_\mu H^\dagger) + h.c.
 \eea
We have checked the $\mathcal{L}^{(5,6,7)}$ matching results with multiple matrix elements to avoid any potential matching ambiguities.
We also note the $N_p$ mass matrix gets perturbed after integrating out $N_1$ or $N_2$. We have determined these corrections,
but as they are dimension eight in the SMEFT they are neglected here.

\section{Perturbation and Non-Perturbation of the neutrino mass matrix}\label{non-pert}
At tree level if the $N_p$ states are integrated out simultaneously or not, the low energy neutrino mass matrix is
perturbed due to $\mathcal{L}^{(7)}$ matchings. The nature of the perturbations are however reflective of the orientations
of the heavy singlet fields in flavour space, as well as their mass spectrum.

It is interesting that a number of effects
that would perturb the low energy neutrino mass matrix cancel out.
For example, the terms in Eqn.~\ref{HEOMterms} cancel in the limit that the Higgs takes on its vacuum expectation value, as previously mentioned.
Integration by parts and EOM manipulations can be used to see this result in the complete basis, when considering the matching onto the
operator $\mathcal{Q}_{\ell H}$. This operator does lead to a contribution to the neutrino mass matrix
when the Higgs takes on its vacuum expectation value
\bea
\langle C_{\ell H}^{\beta \, \kappa} \epsilon_{ij} \epsilon_{mn}(\ell_L^{i \beta} \, C \, \ell_L^{m \kappa}) H^j H^n (H^\dagger H) \rangle =  - \frac{v^4 \, C_{\ell H}^{\beta \, \kappa}}{4} \, (\nu^\beta_L)^T \, \epsilon \, \nu^\kappa_L.
\eea
The dependence on this operator in the expansion of the propagator to third order can be seen to vanish integrated by parts, while
also using Eqn.~\ref{NEOM}. One finds
\bea
\frac{i \, (x_\beta)^T \, x^\star \cdot y^\dagger}{2 \, m_1^3} \, \left[\overline{\ell_{L \, \beta}^{c}} \, \gamma_\mu N_2 \right] \, \tilde{H}^\star \, (H \, D_\mu H^\dagger)
\rightarrow - \frac{(x_\beta)^T \, x^\star \cdot y^\dagger \, y_\delta}{2 \, m_1^3} \,\mathcal{Q}^{\beta \, \delta}_{\ell H} + \cdots
\eea
which cancels the corresponding $\mathcal{Q}_{\ell H}$ term in Eqn.~\ref{dimsevenops}. No additional terms that
contribute to the neutrino mass matrix result from the manipulations in the previous equation; these manipulations also cancel the terms in Eqn.~\ref{L6N23}.
Alternatively, one can integrate out $N_{2,3}$ using the interactions in Eqn.~\ref{L6N23}. Doing so, one finds a contribution to $\mathcal{L}^{(7)}$ that directly cancels the
$\mathcal{Q}_{\ell H}$ dependence in Eqn.~\ref{dimsevenops}. It is important to include $\mathcal{L}^{(6)}_{N_2,N_3}$ and $\mathcal{L}^{(7)}_{N_2,N_3}$ when defining the matching
onto the theory to sub-leading order for this reason. Use of the EOM, and integration by parts on the $N_p$ states still in the spectrum when
$N_1$ is integrated out leads to ambiguities if the full Lagrangian is not specified.

\begin{figure}[h!]
\includegraphics[width=0.51\textwidth]{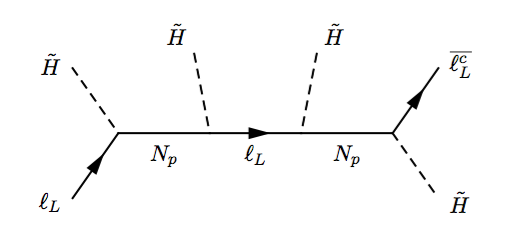}
\caption{Full theory interactions inserted on the tree level propagator to obtain four external $H$ fields.
These scattering contributions only perturb the low energy neutrino mass matrix due to the EOM effect of $\mathcal{L}_5$ modifying Eqn.~\ref{EOMell}.\label{fig3}
Note that when the neutrinos are integrated out in sequence, local contact operators result that lead to even more
mass matrix perturbations, as shown in Fig.\ref{sequent}.}
\end{figure}

The fact that a subset of contributions to $\mathcal{L}^{(7)}$ related to the expansion of the propagator does not
lead to a perturbation of the neutrino mass matrix at tree level
can also been understood intuitively. To obtain $H^\dagger H$  times $\mathcal{Q}_5$ requires
two extra insertions of the coupling of the $N_p$ states to the SM fields. In Eqn.~\ref{basicL} this coupling
is always accompanied by the light SM field $\ell$ so that
no local operator is obtained in the heavy $N_p$ limit expanding the propagator in Feynman diagrams, as illustrated in Fig.\ref{fig3}.
This visual argument is limited, as this fact is not preserved when reducing the operators obtained in the expansion of the propagator by the EOM.
This is another example of the fact that EOM effects in a field theory
do not have a trivial Feynman diagram interpretation.

The detailed nature of the neutrino mass matrix perturbations do change if the states
are integrated out simultaneously or not, as we discuss below.

\subsection{Time ordered products of $\mathcal{L}^{(6)}$ and $\mathcal{L}^{(4)}$}
The limited argument in the previous section also does not forbid perturbations of the neutrino mass matrix due to integrating out the $N_p$ states {\it in sequence}.
Directly expanding out the propagator at tree level to third order, a $\mathcal{L}^{(7)}$ matching contribution comes about due
 to integrating out the heaviest $N_p$ mass eigenstate, and subsequently integrating out the lighter $N_p$ mass eigenstates.
 This always occurs as the $N_p$ cannot be indistinguishable and generate three distinct eigenvalues of the low energy neutrino mass matrix in the UV scenario we consider.
 (Different masses of the $N_p$ states alone still lead to only one low energy eigenvalue of the neutrino mass matrix, only with a different normalization.)
These contributions also match onto $\mathcal{L}^{(7)}$ and lead to additional effects perturbing the neutrino mass matrix.
The action in the EFT generated when the heaviest
$N_p$ state is integrated out has a time ordered product contribution of the form
\bea
S_{eff} = -\frac{1}{2} \int d x^4 \, \int d y^4 \, \mathcal{T}(\mathcal{L}_{N_{2,3}}^6(x), \mathcal{L}^4(y)).
\eea
Reproducing the IR limit in the SMEFT
\begin{figure}[h!]
\includegraphics[width=0.51\textwidth]{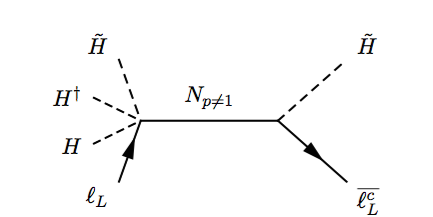}
\caption{Tree level exchange corresponding to the time ordered product $\mathcal{T}(\mathcal{L}^6(x), \mathcal{L}^4(y))$
generating a perturbation to the neutrino mass matrix.\label{sequent}}
\end{figure}
give the following matching contributions
\bea
\mathcal{L}^{(7)} &\supseteq& - \left(\frac{x_\beta^T \, y_\alpha \,  x \cdot y}{2 \, m_1^2 \, m_2}
+ \frac{x_\beta^T \, z_\alpha \, x \cdot z}{2 \, m_1^2 \, m_3} + \frac{y_\beta^T \, z_\alpha  \,  y \cdot z}{2 \, m_2^2 \, m_3}\right) \mathcal{Q}_{\ell H}^{\beta \, \alpha}, \nonumber \\
&\,& - \left(
\frac{x_\beta^T \, y_\alpha \,  y \cdot x}{2 \, m_2^2 \, m_1} + \frac{x_\beta^T \, z_\alpha \,  z \cdot x}{2 \, m_3^2 \, m_1}
+ \frac{y_\beta^T \, z_\alpha \,  z \cdot y}{2 \, m_3^2 \, m_2}
\right) \mathcal{Q}_{\ell H}^{\beta \, \alpha}+ h.c.
\eea
Note that the terms in the second line are generated when
consistently retaining EOM terms (including the $N_p$ that remain in the
spectrum) to reduce the matching contributions to a minimal basis.
Again, for the heavy states to be indistinguishible and integrated out simultaneously,
they would have to be oriented in flavour space in the same manner, and have identical masses. Note that these perturbations to the neutrino mass
matrix are proportional to these differences in flavour space and the multiple mass scales.

\subsection{The neutrino mass matrix up to $\mathcal{L}^{(7)}$}
The contributions to the neutrino mass matrix in the weak eigenbasis up to $\mathcal{L}^{(7)}$ are given by\footnote{Note that the overall sign in these terms is due to a convention choice
on $C$.}
\bea
\mathcal{L}_{\nu \, \nu} &=& \frac{M^{\beta \, \kappa}_{\nu \, \nu}}{2} \nu_L^{T, \, \beta} \, \epsilon \, \nu_L^\kappa, \nonumber \\
 &=&
-\left[\frac{x_\beta^T \, x_\kappa}{m_1} (1+ \frac{v^2 ||x||^2}{2 \, m_1^2}) + \frac{y_\beta^T \, y_\kappa}{m_2}(1+ \frac{v^2 \, ||y||^2}{2 \, m_2^2}) + \frac{z_\beta^T \, z_\kappa}{m_3}(1+ \frac{v^2 \, ||z||^2}{2 \, m_3^2}) \right] \frac{v^2}{4} \nu_L^{T, \, \beta} \, \epsilon \, \nu_L^\kappa, \nonumber \\
&\,& \hspace{0.01cm} -  \left[\frac{x_\beta^T  y_\kappa}{m_1 \, m_2} \, \left(\frac{x \cdot y}{m_1} + \frac{y \cdot x}{m_2} \right) +
\frac{y_\beta^T  z_\kappa}{m_2 m_3}  \, \left(\frac{y \cdot z}{m_2} + \frac{z \cdot y}{m_3} \right)\right] \frac{v^4}{8} \nu_L^{T, \, \beta} \, \epsilon \, \nu_L^\kappa,\\
&\,& \hspace{0.01cm} - \left[\frac{x_\beta^T z_\kappa}{m_1 m_3}  \, \left(\frac{x \cdot z}{m_1} + \frac{z \cdot x}{m_3}\right) \right] \frac{v^4}{8} \nu_L^{T, \, \beta} \, \epsilon \, \nu_L^\kappa
 + h.c. \nonumber
\eea

As the mass matrix is perturbed due to corrections at $\mathcal{L}^{(7)}$ which are suppressed by $\mathcal{O}(v^4/m_p^3)$ {\it and} of order $\mathcal{O}(\omega^4)$.
As such it is established that these corrections can be neglected until perturbations  of the Wilson coefficient in  $\mathcal{L}^{(5)}$
is pushed to relative order $\mathcal{O}(\omega^2 \, v^2/m_p^2)$ compared to leading effects captured by $\mathcal{Q}_5$.
Radiative corrections to $\mathcal{Q}_5$ are generally larger than the non-pertubative corrections due to $\mathcal{L}_7$
and must be incorporated for phenomenological studies as well if these corrections are to be considered.

\section{Flavour Space expansion for the Seesaw}
The expansion that results when integrating out the heavy singlet states in sequence is not the only expansion present in
lower energy Neutrino phenomenology. The usual matching that was developed in the previous sections leads to small perturbations on the
neutrino mass matrix. A larger effect for phenomenology is expanding the Wilson coefficient of the Weinberg operator
systematically due to the perturbations of integrating out the $N_{1,2,3}$ states.
In the remainder of this work, we incorporate and improve on results of Ref.~\cite{Grinstein:2012yu} to develop perturbations of the $\mathcal{U}(\nu,L)$ matricies, assuming a seesaw origin of neutrino mass. We use the SMEFT treatment of the seesaw model developed in the previous sections. The idea is to link perturbations of the PMNS matrix to perturbations of the neutrino mass generation mechanism.\footnote{See also the related (but distinct) sequential dominance idea of S. King discussed in Refs.~\cite{King:1998jw,King:1999cm,King:1999mb,King:2002nf}.} A key point underlying this approach is Majorana mass terms, unlike Dirac mass terms, originate in bi-linears of the same field operators.
As such, the complex mass matrix is diagonalized by a single rotation matrix of the field $\nu_L$ introduced through $\nu_L^p = \mathcal{U}(\nu,L)^{pr} \,  \nu'_{L,r}$.
For this reason any expansion of the neutrino mass matrix is more directly tied to an expansion of the unitary rotation matricies $\mathcal{U}(\nu,L)$.

With the results of the previous section, the Flavour Space Expansion (FSE) of Ref.~\cite{Grinstein:2012yu}  is now on a firmer theoretical footing.
For example,  the heaviest neutrino
in the low energy theory is generically linked to integrating out the lightest singlet field (denoted in this work $N_3$). In Ref.~\cite{Grinstein:2012yu} the neutrino mass matrix is generated by
first integrating out the lightest singlet state, and then integrating out the heavier $N_p$ in sequence. Although this can be done, it is conceptually more clear
to integrate out the three $N_p$ states by removing the heaviest state first, as done here, and subsequently perturb the low energy neutrino mass matrix
after the lighter $N_p$ states are removed in sequence. Doing so the usual SMEFT expansion is present, clarifying the impact of the multiple expansions
present on low energy neutrino phenomenology.

\subsection{Developing the FSE}
The FSE is distinct from the double expansion in $(v/m_p)^n$, and $(E/m_p)^n$ that dictates the relative size of contributions in the SMEFT operator expansion.
This eigenvector perturbation formalism can always be implemented in a type one seesaw model.
However, there is no guarantee that the FSE will be quickly convergent, and therefore predictive,
as it depends upon unknown UV physics parameters.

The basic expectation is that in seesaw models the FSE will be perturbative \cite{Grinstein:2012yu}.
The reason is that the matrix $M_{pr}$ is expected to be approximately uniform in entries in the $N_p$ interaction eigenbasis to the SM states, as the $N_p$ do not carry (known SM) quantum numbers. Diagonalizing the corresponding mass matrix, any hierarchy in the couplings of the $N_R$ states to the SM states is washed out rotating the the
mass eigenbasis, and the magnitude of the $\omega_p$ are drawn together. For this reason it is expected that
\bea\label{generic}
\lVert x \rVert \sim \lVert y \rVert \sim   \lVert z \rVert
\eea
on general grounds.  Here the notation refers to the Euclidean norm of the complex vectors in flavour space. The FSE is of the form
\bea
M^{\beta \, \alpha}_{\nu \, \nu} \, (M^{\kappa \, \alpha}_{\nu \, \nu})^\dagger \simeq  \frac{ \lVert z^\star \cdot z \rVert}{m_3^2} \left[ z_\beta^T \, z_\kappa + \frac{z^\star \cdot y^\dagger}{ \lVert z^\star \cdot z \rVert}  \, \frac{m_3}{m_2} \, z_\beta^T \, y^\star_\kappa
+ \frac{y^\star \cdot z^\dagger}{\lVert z^\star \cdot z \rVert} \, \frac{m_3}{m_2}  \, y_\beta^T \, z^\star_\kappa  + \cdots \right].
\eea
The utility of the FSE depends upon
\bea\label{FSEconvergence}
 \frac{z^\star \cdot y^\dagger}{ \lVert z^\star \cdot z \rVert}   \, \frac{m_3}{m_2} < 1, \quad \quad \frac{y^\star \cdot z^\dagger}{ \lVert z^\star \cdot z \rVert}  \, \frac{m_3}{m_2} <1,
\eea
with similar conditions for integrating out the state of mass $m_1$.  By construction the SMEFT matching has been formulated so that $\frac{m_3}{m_2}  < 1$.
Using the Cauchy-Schwarz equality $a \cdot b = \Delta_{ab} \, \lVert a \rVert \, \lVert b \rVert$ with
 $\Delta_{ab} < 1$ the FSE depends upon
\bea
\frac{\lVert y \rVert}{\lVert z \rVert } \Delta_{y^\dagger z} < m_2/m_3, \quad \quad  \frac{\lVert y \rVert}{\lVert z \rVert } \Delta_{y z^\dagger} < m_2/m_3.
 \eea
Considering Eqn.~\ref{generic}, which directly follows from the quantum numbers of
the $N_p$ states when tuning is avoided, it is expected that the FSE  is present and convergent. In what follows we assume this is the case.
Assuming the FSE exists, the results of Ref.~\cite{Grinstein:2012yu} follow directly, and can be expanded upon in the following way.
To establish notation we define
\bea
M_{\nu \, \nu} = U(\nu,L)^\star \,  {\rm diag}(m_c,m_b,m_a) \, U(\nu,L)^\dagger.
\eea
The rotation matrix is decomposed in eigenvectors such that $U(\nu,L) = \left(\vec{\rho}_c^\star,\vec{\rho}_b^\star,\vec{\rho}_a^\star\right)$ with $\vec{\rho}_i^\star$ a column vector with $\lVert \vec{\rho}_i \rVert = 1$,
and $\lVert \vec{\rho}_i^\star \rVert = 1$.
These eigenvectors are such that
\bea
M_{\nu \, \nu} \, \vec{\rho}_p^\star = m_{p} \, \vec{\rho}_p, \quad \quad m_p > 0, \quad \quad m_p \subset \Re.
\eea
We choose the orthonormal eigenvector basis at leading order to be given by \cite{Grinstein:2012yu}\footnote{See the Appendix for details on the dot and cross products in the flavour space defined over the field $\mathbb{C}^3$.}
\bea\label{LOFSE}
\vec{\rho}_a^\star = \frac{\vec{z}}{\lVert \vec{z} \rVert}, \quad \quad \vec{\rho}_b^\star = \frac{\vec{z}^\star \times (\vec{y} \times \vec{z})}{ \lVert \vec{z} \rVert  \lVert \vec{z} \times \vec{y}\rVert}, \quad \quad
\vec{\rho}_c^\star = \frac{\vec{y}^\star \times \vec{z}^\star}{\lVert \vec{z} \times \vec{y} \rVert}.
\eea
With this convention choice, the mass of the heaviest neutrino is given by $m_a = |\vec{z}|^2 v^2/2 \, m_3$ at leading order and without loss of generality.
The lighter neutrinos are introduced as perturbations \cite{Grinstein:2012yu}. This can also be done without loss of generality.
If the FSE used to introduce these effects is a convergent expansion with small higher order terms depends upon the UV parameters in
the seesaw model.
The perturbations to the eigenvectors and eigenvalues are given in Ref.~\cite{Grinstein:2012yu}. We also define the eigenvectors $U(\nu,L) = \left(\vec{v}_c^\star,\vec{v}_b^\star,\vec{v}_a^\star\right)$ which include the perturbations of the eigenvectors to obtain the full complex mass matrix at dimension five in the SMEFT.
Note that all eigenvectors in this discussion are normalized to ensure unitarity of the PMNS matrix order by order in the
FSE.\footnote{If the PMNS matrix is not unitary, this corresponds to the FSE not converging.}
The PNMS matrix is defined in direct analogy to the CKM matrix as
\bea
\mathcal{U}_{PMNS} = \mathcal{U}^\dagger(e,L) \, \mathcal{U}(\nu, L).
\eea
The rotation matrix $\, \mathcal{U}(e,L)$ is introduced to diagonalize the lepton mass matrix
\bea
\mathcal{M}_e= v \, Y_e/\sqrt{2}, \quad \quad  \mathcal{U}(e,L)^\dagger \, \mathcal{M}^\dagger_e \, \mathcal{M}_e \, \mathcal{U}(e,L) = {\rm diag}\{m_e^2,m_\mu^2,m_\tau^2\}.
\eea
Defining the orthonormal (column) eigenvectors of the lepton rotation matrix $\, \mathcal{U}^\dagger(e,L)$  as $\vec{\sigma}_i$ with $\mathcal{U}^\dagger(e,L)=(\vec{\sigma}_1^\star, \vec{\sigma}_2^\star, \vec{\sigma}_3^\star )^T$
we have
\bea\label{PMNS}
\mathcal{U}^{eigen}_{PNMS} =
\left(\begin{array}{ccc}
\vec{v}_c \cdot \vec{\sigma}_1^\star & \vec{v}_b \cdot \vec{\sigma}_1^\star & \vec{v}_a \cdot \vec{\sigma}_1^\star \\
\vec{v}_c \cdot \vec{\sigma}_2^\star & \vec{v}_b \cdot \vec{\sigma}_2^\star & \vec{v}_a \cdot \vec{\sigma}_2^\star \\
\vec{v}_c \cdot \vec{\sigma}_3^\star & \vec{v}_b \cdot \vec{\sigma}_3^\star & \vec{v}_a \cdot \vec{\sigma}_3^\star \\
\end{array}\right).
\eea
As we are assuming Majorana neutrino masses in a seesaw model, this matrix can be compared to the standard parameterization for unitary matricies.
Define
\bea
P(c_{1},s_1,c_2,s_2,c_3,s_3,\theta) &=&
\left(\begin{array}{ccc}
c_{1} \, c_{3} & s_{1} \, c_{3} & s_{3}  \, e^{- i \theta}\\
- s_{1} \, c_{2}  - c_{1} \, s_{2} \, s_{3} \, e^{i \theta}& c_{1} \, c_{2} - s_{1} \, s_{3} \, s_{2} \, e^{i \theta}& s_{2} \, c_{3} \\
s_{1} \, s_{2} - c_{1} \, c_{2}  \, s_{3} \, e^{i \theta}& -c_{1} \, s_{2} - s_{1} \, c_{2}  \, s_{3} \, e^{i \theta}& c_{2} \, c_{3} \\
\end{array}\right),
\eea
and
\bea
\Theta(v_1,v_2,v_3)= \left(\begin{array}{ccc}
e^{i \, v_{1}} & 0 & 0 \\
0 & e^{i \,v_{2}} & 0 \\
0 & 0& e^{i \, v_{3}}
\end{array}\right),
\eea
so that
\bea
\label{stdparam}
\mathcal{U}^{s_{ij}}_{PNMS} &=&
P(c_{12},s_{12},c_{23},s_{23},c_{13},s_{13},\delta) \Theta(0, \alpha_{21}/2,\alpha_{31}/2),
\eea
with the convention choice $c_{ij} = \cos \theta_{ij}$, $s_{ij} = \sin \theta_{ij}$ and angles $\theta_{ij} = (0,\pi/2)$. Here $\delta = (0,2 \pi)$,
$\alpha_{21}$ and $\alpha_{31}$ are CP violating phases.

Each entry in Eqn.~\ref{PMNS} is a Hermitian inner product characterized by two parameters, naively leading to eighteen parameters.
Comparing to Eqn.~\ref{stdparam} which is a general low energy parameterization in terms of six parameters (three moduli
angles and three phases) makes clear that there is a redundancy of description in this naive interpretation.
However, the eigenvectors sets making up the rotation matricies have to be orthogonal to lead to three masses for the charged leptons and neutrinos.
As such the third vector is not independent in its flavour space orientation, although it can carry a relative phase. This leads to nine parameters in each case.
Using the relation
\bea
\mathcal{U}(\nu, L) = \mathcal{U}(e, L) \, \mathcal{U}^{s_{ij}}_{PNMS},
\eea
with the $\mathcal{U}(\nu, L)$ and $\mathcal{U}(e, L)$ expanded in their eigenvectors we find the leading order result
\begin{align}
\label{solvedrelations}
\vec{\rho}^\star_c &=  (s_{12} \, s_{23} - c_{12} \, c_{23} \, s_{13} \, e^{i \delta}) \,\vec{\sigma}_3
+ (-s_{12} \, c_{23} - c_{12} s_{23}  \, s_{13} e^{i \delta}) \, \vec{\sigma}_2 + c_{12} \, c_{13} \, \vec{\sigma}_1,  \\
\label{solved2}
\vec{\rho}^\star_b \, e^{\frac{- i \alpha_{21}}{2}} &=  \, (- c_{12} \, s_{23} - c_{23} \, s_{12} \, s_{13}  \, e^{i \delta})\,\vec{\sigma}_3
+ (c_{12} \, c_{23} - s_{23} \, s_{12}  \, s_{13} \, e^{i \delta})\,\vec{\sigma}_2
+  c_{13} \, s_{12} \,\vec{\sigma}_1, \\
\label{solvedrelation3}
\vec{\rho}^\star_a \, e^{\frac{-i \alpha_{31}}{2}}&= c_{13} \, c_{23} \,\vec{\sigma}_3
+   c_{13} s_{23} \,\vec{\sigma}_2
+ e^{- i \delta} \, s_{13}  \,\vec{\sigma}_1.
\end{align}
This expression for $\vec{\rho}^\star_a$ then defines $\vec{z}/\lVert \vec{z} \rVert$ at leading order in the FSE.
Further, without loss of generality $\lVert \vec{z} \rVert = 1$ at leading order.
The $\vec{\sigma}_i$ are a set of orthonormal eigenvectors for the unitary matrix $\mathcal{U}(e,L)$.
These vectors form a basis for the field $\mathbb{C}^3$, as they diagonalize $\mathcal{M}_e^\dagger \, \mathcal{M}_e$,  a Hermitian positive matrix also defined over the field $\mathbb{C}^3$.
We can expand the unknown complex flavour vector $y$ into this orthonormal basis. Using the orthogonality and normalization properties of the basis vectors of this space, and a general parameterization of these vectors, then allows the use of the systematic EFT expansion, {\it without the rotation matrix $\mathcal{U}(e,L)$ being chosen to have a fixed form}.
We can always define a flavour vector such that
\bea
\vec{y} = A' \,\vec{\sigma}_1 + B' \,\vec{\sigma}_2 + C' \,\vec{\sigma}_3,
\eea
with $A', B',C' \subset \mathbb{C}$. The vectors $\vec{\sigma}_i$ can be parameterized as discussed in the Appendix.
These vectors satisfy the complex algebra $\vec{\sigma}_i \times \vec{\sigma}_j = \epsilon_{ijk} \, \vec{\sigma}_k$ and $\vec{\sigma}^\star_i \times \vec{\sigma}^\star_j = \epsilon_{ijk} \, \vec{\sigma}^\star_k$
without loss of generality, and we note that the $\vec{\sigma}_i^\star$ are projectable onto $\vec{\sigma}_i$. Solving the general system of equations is straightforward, if tedious.
As an example of the utility of this formalism we examine and falsify a simple case. We show that a UV scenario where the second heavy state integrated out couples to the SM as
\bea\label{simplest}
\vec{y} = A' \,\vec{\sigma}_1,
\eea
does not satisfy Eqns.~\ref{solvedrelations}-\ref{solvedrelation3} and Eqns.~\ref{LOFSE} simultaneously in the limit $s_{13} \rightarrow 0$.\footnote{ As $s_{13}^2 \simeq 0.02$ for $\delta m > 0$ or $\delta m < 0$ \cite{Olive:2016xmw} the limit considered is experimentally motivated.}
This simple example suffices for our purpose of demonstrating how to
perturb in the unknown $\vec{\sigma}_i$ and still obtain physical conclusions on the possible UV theories extending the SM. It is straightforward to derive that in this limit
Eqns.~\ref{solvedrelations}-\ref{solvedrelation3} and Eqns.~\ref{LOFSE} require
\bea
| A' |^2 =\frac{\lVert \vec{z} \times \vec{y} \rVert^2}{c_{13}^2}.
\eea
with the projection coefficients of the $\vec{\sigma}_i^\star$ vectors back onto the $\vec{\sigma}_i$ basis of $\mathbb{C}^3$ required to satisfy (for $s_{13} \rightarrow 0$)
\bea\label{projectinresults}
\sigma_1^\star \cdot \sigma_2 &=& e^{i \, \alpha_{21}/2} \, c_{12} \, c_{23} \,  \frac{\lVert \vec{z} \times \vec{y} \rVert}{(A')^\star}, \\
\sigma_1^\star \cdot \sigma_1 &=& e^{i \, \alpha_{21}/2} \, s_{12} \,  \frac{\lVert \vec{z} \times \vec{y} \rVert}{(A')^\star}, \\
\sigma_3^\star \cdot \sigma_3 &=& \frac{c_{23}^2}{s_{23}^2} \, \sigma_2^\star \cdot \sigma_2 + e^{- i \, \alpha_{31}/2} \, \frac{s_{12}}{s_{23}^2} \, \frac{\lVert \vec{z} \times \vec{y} \rVert}{A'}
- 2 \, e^{- i \, \alpha_{31}/2} \, \frac{c_{23}^2}{s_{23}^2} \, s_{12} \frac{\lVert \vec{z} \times \vec{y} \rVert}{A'}, \\
\sigma_3^\star \cdot \sigma_2  &=& \frac{c_{23}}{s_{23}}  \, \sigma_2^\star \cdot \sigma_2  - e^{-i \, \alpha_{31}/2} \, \frac{c_{23}}{s_{23}} \, s_{12} \,  \frac{\lVert \vec{z} \times \vec{y} \rVert}{A'}, \\
\sigma_3^\star \cdot \sigma_1  &=& \frac{c_{23}}{s_{23}} \, \sigma_2^\star \cdot \sigma_1 + e^{-i \, \alpha_{31}/2} \, \frac{c_{12}}{s_{23}} \,   \frac{\lVert \vec{z} \times \vec{y} \rVert}{A'}, \\
\sigma_2^\star \cdot \sigma_3  &=& \frac{c_{23} \, s_{23}}{c_{23}^2 - s_{23}^2} \, \sigma_3^\star \cdot \sigma_3 - \frac{c_{23} \, s_{23}}{c_{23}^2 - s_{23}^2}  \, \sigma_2^\star \cdot \sigma_2, \\
\sigma_2 \cdot \sigma_3^\star  &=& \frac{c_{23} \, s_{23}}{c_{23}^2 - s_{23}^2}  \, \sigma_3 \cdot \sigma_3^\star - \frac{c_{23} \, s_{23}}{c_{23}^2 - s_{23}^2} \,  \sigma_2 \cdot \sigma_2^\star, \\
\sigma_1^\star \cdot \sigma_3  &=& - \frac{s_{23}}{c_{23}} \sigma_1^\star \cdot \sigma_2, \\
\sigma_1 \cdot \sigma_3^\star  &=& - \frac{s_{23}}{c_{23}} \sigma_1 \cdot \sigma_2^\star.
\eea
This scenario is falsified as it is not possible to simultaneously satisfy these equations using the general parameterization of the unitary matrix
$\mathcal{U}(e,L)$, defining the eigenvectors $\vec{\sigma}_i^\star$ and $\vec{\sigma}_i$. It is easiest to see this point examining the ratio of the
first two equations. The right hand side of this ratio is necessarily $\in \mathbb{R}$, while this does not hold for the left hand side for any
non-zero value of $\delta^\ell$. Vanishing $\delta^\ell$ leads to the other equations not being satisfied.
Note that this conclusion is unchanged if the $\beta_i^\ell$  phases of $\mathcal{U}(e,L)$ are retained or not.
It follows that irrespective of the particular $\vec{\sigma}_i$ chosen, the flavour orientation of the seesaw scenario given in
Eqn.~\ref{simplest} is not consistent at LO in the FSE (for $s_{13} \rightarrow 0$).

\section{Conclusions}
In this paper we have systematically matched the minimal seesaw scenario onto the SMEFT up to dimension seven in the operator
expansion.  We have reported the results on $\mathcal{L}^{(7)}$ in Sec.~\ref{directdim7match}. These corrections can be neglected until
perturbations on the $\mathcal{Q}_5$ operator Wilson coefficient in the Flavour Space Expansion are comparable to a $\omega^2 \, v^2/m_p^2$ SMEFT operator expansion correction. We have shown how the
neutrino mass matrix perturbations due to higher mass dimension operators include effects introduced when integrating out
the $N_p$ states in sequence. We have demonstrated how a consistent matching at $\mathcal{L}^{(6)}$, retaining the $N_p$ in the spectrum after
the $N_1$ state is integrated out, is essential in avoiding matching ambiguities.
We have embedded the Flavour Space Expansion in the SMEFT formalism and  we have developed a novel technique to perturb
in the eigenvectors of the rotation matrix $\mathcal{U}(e,L)$. By treating these vectors as a basis for $\mathbb{C}^3$ to expand the
seesaw theory flavour vectors, one can use the FSE to obtain physical conclusions independent of the form of $\mathcal{U}(e,L)$.
We stress this technique is very general and not limited to the minimal seesaw model, or using the FSE. The results of this work embed the expansions present in neutrino phenomenology
into a well defined effective field theory framework.

\section*{Acknowledgements}
MT acknowledges generous support from the Villum Fonden and partial support by the Danish National Research Foundation (DNRF91). We thank Yun Jiang and Ilaria Brivio for
comments on the manuscript, and an anonymous reviewer for a very good question in the review process on EOM effects.

\appendix
\section{$\mathbb{C}^3$ algebra for eigenvectors diagonalizing a mass matrix}
The dot and cross products act on vectors that have entries defined over the field $\mathbb{C}^3$. The dot product is defined to be a
Hermitian inner product that is anti-linear in its first entry acting on these vectors so that
\bea
x \cdot y = x^\star_i \, y^i,
\eea
with the index $i$ summed over $\{1,2,3\}$. Also note $\lVert x \rVert =\sqrt{x \cdot x}$ and the cross product is defined as
$ x \times y = ((x \times y)_{\Re})^\star$. Here we are indicating complex conjugation of the entries of the usual
cross product defined for vectors, that have entries defined on the field $\Re$. The cross product definition
employed here can actually be formally derived using octonion multiplication \cite{doi:10.1080/14786444508645107,Baez:2001dm},
which also opens up the possibility of further group theory analysis on this approach in flavour space.

Despite the fact that the lepton rotation matrix $\mathcal{U}(e,L)$ is completely unknown, we can perturb around the eigenvectors of this unknown matrix
in the FSE.
The lepton masses are diagonalized in a bi-unitary transformation
\bea
\mathcal{U}(e,R)^\dagger \mathcal{M}_e \, \mathcal{U}(e,L) = {\rm diag} \{m_e \, m_\mu, m_\tau \},
\eea
and $\mathcal{U}(e,L)$ also acts to diagonalize $\mathcal{M}_e^\dagger \, \mathcal{M}_e$. As $\mathcal{U}(e,L)$ is a unitary matrix,
we can parameterize it by the product of three unitary matricies in complete generality
so that
\bea\label{paramel}
\mathcal{U}(e,L)^T = \Theta(\beta^\ell_{1}, \beta^\ell_{2},\beta^\ell_{3})
 \, P(c^\ell_{12},s^\ell_{12},c^\ell_{23},s^\ell_{23},c_{13}^\ell,s_{13}^\ell,\delta^\ell)
 \Theta(\alpha^\ell_{1}, \alpha^\ell_{2},\alpha^\ell_{3}).
\eea
This introduces ten parameters into the parameterization of this matrix, instead of the usual nine parameters for a unitary $3 \times 3$ matrix.
We use the redundancy in one phase introduced to establish the algebra of the eigenvectors we wish to perturb in.
From this general parameterization we have
\bea
\frac{\sigma_1^T}{e^{i \beta^\ell_1}} &=& \left\{c^\ell_{12} \, c^\ell_{13} e^{i \, \alpha^\ell_1}, c^\ell_{13} \, s^\ell_{12} e^{i \, \alpha_2^\ell}, s^\ell_{13} \, e^{i
(\alpha_3^\ell  - \delta^\ell)}\right\}, \\
\frac{\sigma_2^T}{e^{i \beta^\ell_2}}  &=&  \left\{\left(-c_{23}^\ell \, s_{12}^\ell - c_{12}^\ell  \, s^\ell_{13} \, s^\ell_{23} \, e^{i \delta^\ell} \right) e^{i \alpha_1^\ell},
 \left(c^\ell_{12} \, c^\ell_{23} - s^\ell_{12} \, s^\ell_{13} s_{23}^\ell e^{i \delta^\ell} \right) e^{i \alpha_2^\ell}, c_{13}^\ell \, s_{23}^\ell \, e^{i \alpha_3^\ell}\right\},  \\
\frac{\sigma_3^T}{e^{i \beta^\ell_3}}  &=& \left\{\left(s_{23}^\ell \, s_{12}^\ell - c_{12}^\ell  \, s^\ell_{13} \, c^\ell_{23} \, e^{i \delta^\ell} \right) e^{i \alpha_1^\ell},
 \left(-c^\ell_{12} \, s^\ell_{23} - s^\ell_{12} \, s^\ell_{13} c_{23}^\ell e^{i \delta^\ell} \right) e^{i \alpha_2^\ell}, c_{13}^\ell \, c_{23}^\ell \, e^{i \alpha_3^\ell}\right\}.
\eea
One can then directly determine the complex algebra $\vec{\sigma}_i \times \vec{\sigma}_j = \epsilon_{ijk} \, \vec{\sigma}_k$ is present
when the phase convention choice $\alpha^\ell_1+ \alpha^\ell_2 + \alpha^\ell_3+ \beta^\ell_1+ \beta^\ell_2 + \beta^\ell_3 = 2 \, \pi \, n$ , $n \subset \mathbb{Z}$
is made. A phase choice of this form is allowed, and reduces the number of free parameters in the parameterization of the unitary matrix $\mathcal{U}(e,L)$ to nine.
It follows directly that $\vec{\sigma}^\star_i \times \vec{\sigma}^\star_j = \epsilon_{ijk} \, \vec{\sigma}^\star_k$ in general.
It is also required to know the projection coefficients of the $\vec{\sigma}_i^\star$ onto the basis of vectors $\vec{\sigma}_i$ to perform the eigenvector
perturbations in a general way. They can be derived directly using the definition of the Hermitian inner product and recalling
$\vec{\sigma}_i^\star \cdot \vec{\sigma}_j =(\vec{\sigma}_j \cdot \vec{\sigma}^\star_i)^\star$.

To simplify the intermediate steps of the calculation involving $\vec{\sigma}_i$
and $\vec{\sigma}_i^\star$ it can be convenient to re-phase the charged lepton field to make
the eigenvalues of $\mathcal{M}_e^\dagger \, \mathcal{M}_e$ positive and remove the $\alpha_i$ from Eqn.~\ref{paramel} without physical effect.\footnote{For related discussions see Refs.\cite{Jenkins:2007ip,Jenkins:2009dy,Jarlskog:1985cw,Kusenko:1993ph}.}
The $\beta_i$ phases in Eqn.~\ref{paramel} define a similarity transformation
\bea\label{similarity}
\mathcal{M}_e^\dagger \, \mathcal{M}_e \rightarrow \Theta(\beta^\ell_{1}, \beta^\ell_{2},\beta^\ell_{3})^\dagger \, \mathcal{M}_e^\dagger \, \mathcal{M}_e \,  \Theta(\beta^\ell_{1}, \beta^\ell_{2},\beta^\ell_{3}).
\eea
One can also choose a parameterization of $\mathcal{U}(e,L)$ where these $\beta_i$ intermediate unphysical phases vanish.
This similarity transformation leaves the eigenvalues of $\mathcal{M}_e^\dagger \, \mathcal{M}_e$ invariant
but does not leave the eigenvectors invariant in general. Choosing this phase convention fixes a general class of $\vec{\sigma}_i$ to perturb around as a basis for $\mathbb{C}^3$.
As  physically observable effects due to $\mathcal{U}_{PMNS}$ only come about due to the relationship between the eigenvectors sets $\vec{\sigma}_i^\star$ and $\rho_i^\star$
this can be done as a convention choice.
\section{Operator basis of Ref.~\cite{Lehman:2014jma}}
\begin{table}[h!]
\begin{center}
\small
\begin{minipage}[t]{8.cm}
\renewcommand{\arraystretch}{1.5}
\begin{tabular}[t]{c|c}
\multicolumn{2}{c}{$1: \psi^2 H^4 + \hbox{h.c.}$} \\
\hline
$\mathcal{Q}_{\ell H}$                & $\epsilon_{ij} \epsilon_{mn}(\ell_L^i C \ell_L^m) H^j H^n (H^\dagger H)$
\end{tabular}
\end{minipage}
%
%
%
%
%
%
\begin{minipage}[t]{6.5cm}

\renewcommand{\arraystretch}{1.5}
\begin{tabular}[t]{c|c}
\multicolumn{2}{c}{$2: \psi^2 H^2 D^2 + \hbox{h.c.}$} \\
\hline
$\mathcal{Q}_{\ell H D}^{(1)}$           & $\epsilon_{ij} \epsilon_{mn}\ell_L^i C (D^\mu \ell_L^j)H^m (D_\mu H^n)$ \\
$\mathcal{Q}_{\ell H D}^{(2)}$          & $\epsilon_{im} \epsilon_{jn}\ell_L^i C (D^\mu \ell_L^j)H^m (D_\mu H^n)$
\end{tabular}
\end{minipage}

\vspace{0.25cm}

\begin{minipage}[t]{8.cm}
\renewcommand{\arraystretch}{1.5}
\begin{tabular}[t]{c|c}
\multicolumn{2}{c}{$3: \psi^2 H^3 D + \hbox{h.c.}$} \\
\hline
$\mathcal{Q}_{\ell H D e}$     & $\epsilon_{ij} \epsilon_{mn} (\ell_L^i C \gamma_\mu e_R) H^j H^m D^\mu H^n$
\end{tabular}
\end{minipage}
%
%
%
%
\begin{minipage}[t]{6.5cm}
\renewcommand{\arraystretch}{1.5}
\begin{tabular}[t]{c|c}
\multicolumn{2}{c}{$4:\psi^2 H^2 X+ \hbox{h.c.}$} \\
\hline
$\mathcal{Q}_{\ell H B}$      & $\epsilon_{ij} \epsilon_{mn} (\ell_L^i C \sigma_{\mu \nu} \ell_L^m) H^j H^n B^{\mu \nu}$\\
$\mathcal{Q}_{\ell H W}$      & $\epsilon_{ij} (\tau^I \epsilon)_{mn} (\ell_L^i C \sigma_{\mu \nu} \ell_L^m) H^j H^n W^{I \mu \nu} $
\end{tabular}
\end{minipage}

\vspace{0.25cm}

\begin{minipage}[t]{8.cm}
\renewcommand{\arraystretch}{1.5}
\begin{tabular}[t]{c|c}
\multicolumn{2}{c}{$5:\psi^4 D + \hbox{h.c.}$} \\
\hline
$\mathcal{Q}_{\ell \ell \overline{d} u D}^{(1)}$        & $\epsilon_{ij} (\overline{d_R} \gamma_\mu u_R)(\ell_L^i C D^\mu \ell_L^j )$ \\
$\mathcal{Q}_{\ell \ell \overline{d} u D}^{(2)}$        & $\epsilon_{ij} (\overline{d_R} \gamma_\mu u_R)(\ell_L^i C \sigma^{\mu \nu} D_\nu \ell_L^j )$ \\
$\mathcal{Q}_{\overline{\ell} Q d d D}^{(1)}$  & $(Q_L C \gamma_\mu d_R)(\overline{\ell_L} D^\mu d_R)$ \\
$\mathcal{Q}_{\overline{\ell} Q d d D}^{(2)}$  & $(\overline{\ell_L} \gamma_\mu q_L)(d_R C D^\mu d_R)$ \\
$\mathcal{Q}_{d d d \overline{e} D}$                & $(\overline{e_R} \gamma_\mu d_R)(d_R C D^\mu d_R)$
\end{tabular}
\end{minipage}
\begin{minipage}[t]{6.5cm}
\renewcommand{\arraystretch}{1.5}
\begin{tabular}[t]{c|c}
\multicolumn{2}{c}{$6:\psi^4 H + \hbox{h.c.}$} \\
\hline
$\mathcal{Q}_{\ell \ell \ell \overline{e} H}$               & $\epsilon_{ij} \epsilon_{mn} (\overline{e_R} \ell_L^i)(\ell_L^j C \ell_L^m ) H^n$ \\
$\mathcal{Q}_{\ell \ell Q \overline{d} H}^{(1)}$        & $\epsilon_{ij} \epsilon_{mn} (\overline{d_R} \ell_L^i)(q_L^j C \ell_L^m ) H^n$ \\
$\mathcal{Q}_{\ell \ell Q \overline{d} H}^{(2)}$        & $\epsilon_{im} \epsilon_{jn} (\overline{d_R} \ell_L^i)(q_L^j C \ell_L^m ) H^n$ \\
$\mathcal{Q}_{\ell \ell \overline{Q} u H}$                      & $\epsilon_{ij}(\overline{q_{L_m}} u_R )(\ell_L^m C \ell_L^i)H^j$ \\
$\mathcal{Q}_{\overline{\ell} Q Q d H}$                      & $\epsilon_{ij}(\overline{\ell_{L_m}} d_R)(q_L^m C q_L^i) \tilde{H}^j$ \\
$\mathcal{Q}_{\overline{\ell} d d d H}$                & $(d_R C d_R)(\overline{\ell_L} d_R)H$ \\
$\mathcal{Q}_{\overline{\ell} u d d H}$                & $(\overline{\ell_L} d_R)(u_R C d_R) \tilde{H}$ \\
$\mathcal{Q}_{\ell e u \overline{d} H}$                & $\epsilon_{ij} (\ell_L^i C \gamma_\mu e_R) (\overline{d_R} \gamma^\mu u_R) H^j$ \\
$\mathcal{Q}_{\overline{e} Q d d H}$                & $\epsilon_{ij} (\overline{e_R} Q_L^i)(d_R C d_R )\tilde{H}^j$ \\
\end{tabular}
\end{minipage}
%
%
%
%
\end{center}
\caption{\label{LehmanBas}
The operator basis of Ref.~\cite{Lehman:2014jma} matched onto in this work. Here the spinors are in four component notation
and $C= - i \gamma_2 \, \gamma_0$ in the chiral basis we employ.}
\end{table}
\newpage
\bibliographystyle{JHEP}
\bibliography{bibliography2}
\end{document}